\documentclass[showpacs,aps,prb,twocolumn,amssymb]{revtex4}
\usepackage{graphicx}
\usepackage{amsmath}

\begin{document}

\title{Stable Sarma State in Two-band Fermi Systems}
\author{Lianyi He and Pengfei Zhuang}

\affiliation{Physics Department, Tsinghua University, Beijing
100084, China}

\begin{abstract}
We investigate fermionic superconductivity with mismatched Fermi
surfaces in a general two-band system. The exchange interaction
between the two bands changes significantly the stability
structure of the pairing states. The Sarma state with two gapless
Fermi surfaces which is always unstable in single-band systems,
can be the stable ground state in two-band systems. To realize a
visible mismatch window for the stable Sarma state, two conditions
should be satisfied: a nonzero inter-band exchange interaction and
a large asymmetry between the two bands.
\end{abstract}

\maketitle

\section {Introduction} \label{s1}
The Cooper pairing with mismatched Fermi surfaces, which has been
investigated many years ago~\cite{sarma,LOFF}, promoted new
interest in the study of new superconducting materials in strong
magnetic field and ultracold fermions due to the realization of
superfluidity in resonantly interacting Fermi gases. The
well-known theoretical result for s-wave weak coupling
superconductors is that, at a critical mismatch, called
Chandrasekhar-Clogston limit (CC limit) $h_c=0.707\Delta_0$ where
$\Delta_0$ is the zero temperature gap, a first order phase
transition from the gapped BCS state to the normal state
occurs~\cite{CC}. Further theoretical studies showed that the
inhomogeneous Fulde-Ferrell-Larkin-Ovchinnikov (FFLO)
state~\cite{LOFF} may survive in a narrow window between $h_c$ and
$h_{\text{FFLO}}=0.754\Delta_0$. However, since the thermodynamic
critical field is much smaller than the CC limit due to strong
orbit effect~\cite{CC}, it is hard to observe the CC limit and the
FFLO state in ordinary superconductors. In recent years, some
experimental evidence for the FFLO state in heavy fermion
superconductors~\cite{heavy}, high temperature
supercondutors~\cite{HTSC} and organic
superconductors~\cite{organic} were found. More recently, in the
study of ultracold atoms, Fermi superfluidity with population
imbalance was realized by MIT and Rice groups
independently~\cite{exp}. The ultracold fermion experiments has
promoted a lot of theoretical
works~\cite{BCS,crossover,trap,review} on the superfluidity
mechanism and the phase diagrams for the crossover from
Bardeen-Cooper-Schrieffer (BCS) to Bose-Einstein Condensation(BEC)
~\cite{BCSBEC}. The problem of imbalanced pairing is also related
to the study of color superconductivity and pion superfluidity in
dense quark matter~\cite{CSC,pion}.

While most of the theoretical works focus on the inhomogeneous
FFLO state, we in this paper are interested in the homogeneous and
gapless Sarma state~\cite{sarma}. For weak coupling
superconductors, the Sarma state is located at the maximum of the
thermodynamic potential of the system, and therefore can not be
the stable ground state. This was called Sarma instability many
years ago~\cite{sarma}. The thermodynamic instability of the Sarma
state can be traced to the existence of gapless fermion
excitations which cause a very large density of state at the
gapless Fermi surfaces~\cite{sarma,LOFF}. To realize a stable
Sarma state, one should have some mechanism to cure the
instability. Forbes et.al.~\cite{BP} proposed that, a stable Sarma
state is possible in a model with finite range interaction where
the momentum dependence of the pairing gap cures the instability.
On the other hand, when the attractive interaction becomes strong
enough which can be realized in ultracold fermion experiments, the
stability of the Sarma state can be changed. While the homogeneous
Sarma state is always unstable at the BCS side of the BCS-BEC
crossover, it becomes stable in the deep BEC
region~\cite{crossover}. However, this stable Sarma state at the
BEC side is not the original ``interior gap" or ``breached
pairing" state with two gapless Fermi surfaces proposed by Liu and
Wilczek~\cite{IG}. Since the fermion chemical potential becomes
negative in the BEC region, the Sarma state in this case possesses
only one gapless Fermi surface, and the matter behaves like a
Bose-Fermi mixture~\cite{BFM}.

In this paper, we focus on how the multi-band structure which may
be realized in solid materials and optical lattices changes the
stability of the Sarma state. We consider a general two-band Fermi
system, and show that the inter-band exchange interaction can cure
the Sarma instability and the Sarma state can be the stable ground
state in visible parameter regions.

The multi-band theory of BCS superconductivity was firstly
introduced by Suhl et al.~\cite{twoband} in 1959 to describe the
possible multiple band crossings at the Fermi surface. The
two-band model has been applied to the study of high-$T_c$
superconductors~\cite{Loktev} to effectively describe the
particular crystalline and electronic structure. Recently, it is
found that, the material $\rm {MgB_2}$ is a standard two-band
superconductor~\cite{mgb} and many experimental data can be
explained by the two-band model of BCS superconductivity.
Multi-band Fermi systems may be realized experimentally with
ultracold atoms in optical lattice~\cite{kohl}. For example, if we
confine the cold atoms in a one dimensional periodic external
potential, the band structure will form in the confined direction,
and the matter can be regarded as a multi-band system in
two-dimensions. In this case, by adjusting the coupling strength,
one can study the possible BCS-BEC crossover in multi-band
systems~\cite{Loktev,iskin}. The inter-band physics in optical
lattices is recently studied~\cite{inter}, and the multi-gap
superfluidity is also possible in nuclear matter~\cite{nuclear}.

The paper is organized as follows. In Section \ref{s2} we give an
introduction to the Sarma state in single-band systems. We discuss
the stability of the Sarma state in a general two-band model in
Section \ref{s3} and summarize in Section \ref{s4}.

\section {Sarma State in Single-band Model} \label{s2}
Before discussing the stability of Sarma state in two-band
systems, we in this section give a brief introduction to the Sarma
state in single-band systems. We start from the following standard
BCS-type Hamiltonian
\begin{eqnarray}
H&=&\int d^3{\bf r}\bigg[\sum_{\sigma}\psi^\dagger_{\sigma}({\bf
r})\left(-{\nabla^2\over 2m}-\mu_{\sigma}\right)\psi^{\phantom{\dag}}_{\sigma}({\bf r})\nonumber\\
&-&U^{\phantom{\dag}}\psi^\dagger_{\uparrow}({\bf
r})\psi^\dagger_{\downarrow}({\bf
r})\psi^{\phantom{\dag}}_{\downarrow}({\bf
r})\psi^{\phantom{\dag}}_{\uparrow}({\bf r})\bigg].
\end{eqnarray}
We constrain ourselves to discuss systems at zero temperature
where the BCS mean field theory can be applied even at strong
coupling~\cite{BCSBEC}. In the mean field approximation, the
Hamiltonian is approximated by
\begin{eqnarray}
H_{\text{mf}}&=&\int d^3{\bf
r}\Bigg[\sum_{\sigma}\psi^\dagger_{\sigma}({\bf r})
\left(-\frac{\nabla^2}{2m}-\mu_{\sigma}\right)\psi^{\phantom{\dag}}_{\sigma}({\bf r})\nonumber\\
&+&\Phi({\bf r})\psi^\dagger_{\uparrow}({\bf
r})\psi^\dagger_{\downarrow}({\bf r})+\text{H.c.}+\frac{|\Phi({\bf
r})|^2}{U}\Bigg],
\end{eqnarray}
where $\Phi({\bf r})=-U\langle
\psi^{\phantom{\dag}}_{\downarrow}({\bf
r})\psi^{\phantom{\dag}}_{\uparrow}({\bf r})\rangle$ is the order
parameter field of superconductivity. For homogeneous
superconductivity, the thermodynamic potential $\Omega$ can be
obtained by using the standard diagonal method~\cite{twoband}. It
can be expressed as
\begin{eqnarray}
\Omega=\frac{\Delta^2}{U}+\int{d^3{\bf k}\over
(2\pi)^3}\bigg[(\xi_{{\bf k}}-E_{{\bf k}})+\sum_\sigma E_{{\bf
k}}^\sigma\Theta(-E_{{\bf k}}^\sigma)\bigg]
\end{eqnarray}
with the definition of energy dispersions $\xi_{{\bf k}}={\bf
k}^2/(2m)-\mu$, $E_{{\bf k}}=\sqrt{\xi_{{\bf k}}^2+\Delta^2}$,
$E_{{\bf k}}^\uparrow=E_{{\bf k}}+ h$ and $E_{{\bf
k}}^\downarrow=E_{{\bf k}}- h$, where
$\mu=(\mu_{\uparrow}+\mu_{\downarrow})/2$ and
$h=(\mu_{\uparrow}-\mu_{\downarrow})/2$ are, respectively, the
averaged and mismatched chemical potentials, and $\Delta$ is the
modulus of $\Phi({\bf r})$.

Without loss of generality, we set $h\geq 0$. The possible ground
state of the system corresponds to the stationary point of the
thermodynamic potential $\Omega$. This gives the so-called gap
equation
\begin{equation}
\left(\frac{1}{U}-\int{d^3{\bf k}\over
(2\pi)^3}\frac{\Theta(E_{{\bf k}}^\downarrow)}{2E_{{\bf
k}}}\right)\Delta=0.
\end{equation}
To properly achieve strong coupling, the chemical potentials
should be renormalized by the number equations. The number density
$n$ and spin density imbalance $\delta$ can be evaluated as
\begin{eqnarray}
n &=& n_{\uparrow}+n_{\downarrow}=\int{d^3{\bf k}\over
(2\pi)^3}\left[1-\frac{\xi_{{\bf k}}}{E_{{\bf
k}}}\Theta(E_{{\bf k}}^\downarrow)\right],\nonumber\\
\delta &=& n_{\uparrow}-n_{\downarrow}=\int{d^3{\bf k}\over
(2\pi)^3}\Theta(-E_{{\bf k}}^\downarrow).
\end{eqnarray}
Whether the Zeeman energy imbalance $h$ or the spin density
imbalance $\delta$ is experimentally adjusted depends on detailed
systems. For cold atoms, the spin density imbalance $\delta$ is
directly tuned, but in superconductors, the Zeeman splitting $h$
is adjusted via an external magnetic field.

\subsection {Stability Analysis}
If a solution of the gap equation is the ground state of the
system, it should be the global minimum of the thermodynamic
potential $\Omega$~\cite{BP,comment}. The condition for a local
minimum of $\Omega$ is that
\begin{equation}
\frac{\partial\Omega(\Delta)}{\partial\Delta}=0,\ \ \
\frac{\partial^2\Omega(\Delta)}{\partial\Delta^2}>0.
\end{equation}
The first condition corresponds to the gap equation and the second
order derivative $I=\partial^2\Omega(\Delta)/\partial\Delta^2$ can
be evaluated as
\begin{equation}
I=\int{d^3{\bf k}\over (2\pi)^3}\frac{\Delta^2}{E_{{\bf
k}}^2}\bigg[\frac{\Theta(E_{{\bf k}}^\downarrow)}{E_{{\bf
k}}}-\delta(E_{{\bf k}}^\downarrow)\bigg].
\end{equation}

Let us study the Sarma state with $h>\Delta$ which induces a
nonzero spin density imbalance $\delta$. At weak coupling, $I$ can
be approximately evaluated as
\begin{equation}
\frac{\pi^2I}{m}\simeq\sqrt{2m\mu}\left[1-\frac{h\Theta(h-\Delta)}{\sqrt{h^2-\Delta^2}}\right]
\end{equation}
which shows that $\partial^2\Omega(\Delta)/\partial\Delta^2$ is
always negative and therefore the Sarma state is unstable.

To achieve the BCS-BEC crossover, we renormalize the coupling
constant with the two-body scattering length $a_s$,
\begin{equation}
\frac{m}{4\pi a_s}=-\frac{1}{U}+\int\frac{d^3{\bf
k}}{(2\pi)^3}\frac{m}{{\bf k}^2}.
\end{equation}
In this case, we first solve the coupled gap and number equations
at fixed total density $n=k_{\text F}^3/(3\pi^2)$. The result can
be expressed~\cite{crossover} as a function of the dimensionless
coupling parameter $g=1/(k_{\text F}a_s)$ and the population
imbalance $P=\delta/n$. The numerical calculations show that, the
key quantity $I$ is always negative at the BCS side of the
resonance($a_s<0, g<0$) where the Sarma state has two gapless
Fermi surfaces, but the Sarma state can be a stable ground state
in the strong coupling BEC region (roughly for $g>2.2$) where the
chemical potential $\mu$ become negative. However, this Sarma
state has only one gapless Fermi surface and is different from the
Fermi surface topology of the so-called breached pairing state.

\subsection {Solution at Weak Coupling}
At weak coupling where the chemical potential $\mu$ is well
approximated by the Fermi energy $\epsilon_{\text F}$, the gap
equation can be approximated by
\begin{equation}
\left[\frac{1}{UN}-\int_0^\Lambda
d\xi\frac{\Theta(\sqrt{\xi^2+\Delta^2}-h)}{\sqrt{\xi^2+\Delta^2}}\right]\Delta=0,
\end{equation}
where $N$ is the density of state for each spin state at the Fermi
surface, and $\Lambda$ is the energy cutoff which plays the role
of Debye energy $\hbar\omega_{\text D}$ in solids. After the
integration and using the condition $\Delta\ll\Lambda$, we find
\begin{equation}
\left[\frac{1}{UN}-\ln\frac{2\Lambda}{\Delta}+\Theta(h-\Delta)\ln\frac{h+\sqrt{h^2-\Delta^2}}{\Delta}\right]\Delta=0.\label{gap1}
\end{equation}

There are three possible solutions to the gap equation
(\ref{gap1}) for $h\neq 0$. The first is the trivial normal phase
with $\Delta_{\text N}=0$. The second corresponds to the ordinary
fully gapped BCS solution satisfying $\Delta>h$,
\begin{equation}
\Delta_{\text{BCS}}=\Delta_0=2\Lambda e^{-1/(UN)}.
\end{equation}
The third solution, i.e., the gapless Sarma state satisfying
$\Delta<h$, can be analytically evaluated via the comparing with
the BCS solution. It is~\cite{Nardulli}
\begin{eqnarray}
\Delta_{\text S}=\sqrt{\Delta_0(2h-\Delta_0)}.
\end{eqnarray}

Using the weak coupling approximation, the grand potential
$\Omega$ for various solutions can be expressed as
\begin{eqnarray}
\Omega&=&\frac{\Delta^2}{U}+2N\int_0^\Lambda
d\xi\Big[\xi-\sqrt{\xi^2+\Delta^2}\nonumber\\
&+&(\sqrt{\xi^2+\Delta^2}-h)\Theta(h-\sqrt{\xi^2+\Delta^2})\Big].
\end{eqnarray}
Performing the integral over $\xi$, and using the condition
$\Delta\ll\Lambda$ as well as the gap equation to cancel the
cutoff dependence, we have
\begin{eqnarray}
\Omega=-\frac{N}{2}\Delta^2-\Theta(h-\Delta)Nh\sqrt{h^2-\Delta^2}.
\end{eqnarray}
Note that we have set the grand potential of the normal state at
$h=0$ to be zero, $\Omega_{\text N}(h=0)=0$. To see why the Sarma
state is always thermodynamically unstable, one should calculate
the grand potential differences between Sarma and other two
states~\cite{Nardulli},
\begin{eqnarray}
\Omega_{\text S}-\Omega_{\text
{BCS}}&=&N(\Delta_0-h)^2,\nonumber\\
\Omega_{\text{S}}-\Omega_{\text N}&=&\frac{N}{2}(\Delta_0-2h)^2,
\end{eqnarray}
which confirm that the Sarma state always has higher potential
than the BCS and normal states. As a consequence, there exists a
first order phase transition from BCS state to normal state. From
the result
\begin{eqnarray}
\Omega_{\text{BCS}}-\Omega_{\text N}=\frac{N}{2}(2h^2-\Delta_0^2),
\end{eqnarray}
the transition occurs at the CC limit of BCS superconductivity,
$h_c=\Delta_0/\sqrt{2}$.

\section {Sarma State in Two-band Model} \label{s3}
We in this section turn to the two-band model. Since the goal of
this paper is to search for the possibility of stable Sarma state
in general two-band Fermi systems, we consider a continuum
Hamiltonian and neglect the details of the band structure in
different systems. We will show that the key point is the
inter-band scattering which can make the Sarma state stable in
multi-band systems. The possible complicated lattice structure in
various materials and optical lattices will not qualitatively
change our conclusion. The obtained conclusion is generic and may
be useful for the study of superconducting materials and ultracold
atom gases.

The continuum Hamiltonian of the two-band model can be written
as~\cite{twoband}
\begin{eqnarray}
H&=&\int d^3{\bf
r}\bigg[\sum_{\nu,\sigma}\psi^\dagger_{\nu\sigma}({\bf
r})\left(-{\nabla^2\over 2m_\nu}-\mu_{\nu\sigma}\right)\psi^{\phantom{\dag}}_{\nu\sigma}({\bf r})\nonumber\\
&-&\sum_{\nu,\lambda}U^{\phantom{\dag}}_{\nu\lambda}\psi^\dagger_{\nu\uparrow}({\bf
r})\psi^\dagger_{\nu\downarrow}({\bf
r})\psi^{\phantom{\dag}}_{\lambda\downarrow}({\bf
r})\psi^{\phantom{\dag}}_{\lambda\uparrow}({\bf r})\bigg],
\end{eqnarray}
where $\nu,\lambda=1,2$ denote the band and
$\sigma=\uparrow,\downarrow$ the direction of fermion spin. In
superconductors, the band degrees of freedom usually come from the
particular crystalline and electronic structure of the materials.
In ultracold atom gases, these degrees of freedom may come from
different hyperfine states or different atom species or the
external periodic lattice potential. In general case, the
effective fermion mass depends only on the band index, but the
chemical potential is related to both the band and spin indexes
due to the existence of external magnetic field or population
imbalance. The constants $U_{11}\equiv U_1$ and $U_{22}\equiv U_2$
are the intra-band couplings, and $U_{12}=U_{21}\equiv J$ is the
inter-band exchange coupling. For vanishing $J$, the model is
reduced to a simple system with two independent bands. In the
following we focus on how the inter-band coupling $J$ changes the
stability of the Sarma state.

We first calculate the thermodynamic potential of the two-band
Hamiltonian. In the mean field approximation, the Hamiltonian is
approximated by
\begin{eqnarray}
H_{\text{mf}}&=&\int d^3{\bf
r}\bigg\{\sum_{\nu,\sigma}\psi^\dagger_{\nu\sigma}({\bf r})
\left(-\frac{\nabla^2}{2m_\nu}-\mu_{\nu\sigma}\right)\psi^{\phantom{\dag}}_{\nu\sigma}({\bf r})\nonumber\\
&+&\sum_{\nu}\left[\Phi_\nu({\bf
r})\psi^\dagger_{\nu\uparrow}({\bf
r})\psi^\dagger_{\nu\downarrow}({\bf
r})+\text{H.c.}\right]\nonumber\\
&+&\frac{1}{G}\Big[U_{2}|\Phi_1({\bf r})|^2+U_{1}|\Phi_2({\bf
r})|^2\nonumber\\
&-&J\left(\Phi_1^*({\bf r})\Phi_2({\bf r})+\Phi_2^*({\bf
r})\Phi_1({\bf r})\right)\Big]\bigg\},
\end{eqnarray}
where $\Phi_\nu({\bf
r})=-\sum_{\lambda}U^{\phantom{\dag}}_{\nu\lambda}\langle
\psi^{\phantom{\dag}}_{\lambda\downarrow}({\bf
r})\psi^{\phantom{\dag}}_{\lambda\uparrow}({\bf r})\rangle$ are
two order parameter fields of the superconductivity, and $G$ is
defined as $G=U_{1}U_{2}-J^2$. For homogeneous superconductivity,
the thermodynamic potential $\Omega$ of this two-band model can
be obtained by using the standard diagonal method~\cite{twoband}.
It can be expressed as
\begin{eqnarray}
\Omega&=&\frac{1}{G}\left[U_{2}\Delta_1^2+U_{1}\Delta_2^2-2J\Delta_1\Delta_2\cos(\varphi_1-\varphi_2)\right]\\
&+&\sum_\nu\int{d^3{\bf k}\over (2\pi)^3}\bigg[(\xi_{{\bf
k}\nu}-E_{{\bf k}\nu})+\sum_\sigma E_{{\bf
k}\nu}^\sigma\Theta(-E_{{\bf k}\nu}^\sigma)\bigg]\nonumber
\end{eqnarray}
with the definition of energy dispersions $\xi_{{\bf k}\nu}={\bf
k}^2/(2m_\nu)-\mu_\nu$, $E_{{\bf k}\nu}=\sqrt{\xi_{{\bf
k}\nu}^2+\Delta_\nu^2}$, $E_{{\bf k}\nu}^\uparrow=E_{{\bf k}\nu}+
h_\nu$ and $E_{{\bf k}\nu}^\downarrow=E_{{\bf k}\nu}- h_\nu$,
where $\mu_\nu=(\mu_{\nu\uparrow}+\mu_{\nu\downarrow})/2$ and
$h_\nu=(\mu_{\nu\uparrow}-\mu_{\nu\downarrow})/2$ are,
respectively, the averaged and mismatched chemical potentials, and
$\Delta_\nu$ the modulus of $\Phi_\nu$ and $\varphi_\nu$ their
phases through the definition $\Phi_\nu=\Delta_\nu
e^{i\varphi_\nu}$. Without loss of generality, we take $h_\nu >
0$. For $J>0$, the choice of $\varphi_1=\varphi_2$ is favored,
otherwise there is $\varphi_1=\varphi_2+\pi$. We assume $J>0$ and
set $\varphi_1=\varphi_2$.

The possible ground state of the system corresponds to the
stationary point of the thermodynamic potential $\Omega$. This
gives the so-called gap equations
\begin{equation}
\left(\frac{U_{\bar\nu\bar\nu}}{G}-\int{d^3{\bf k}\over
(2\pi)^3}\frac{\Theta(E_{{\bf k}\nu}^\downarrow)}{2E_{{\bf
k}\nu}}\right)\Delta_\nu-\frac{J}{G}\Delta_{\bar\nu}=0
\end{equation}
with $\bar\nu=1$ for $\nu=2$ and $\bar\nu=2$ for $\nu=1$. The gap
equations are essentially the same as derived in \cite{twoband}.
To properly achieve strong coupling, the chemical potentials
should be renormalized by the number equations. The number
equations for the fermion density $n_\nu$ and density imbalance
$\delta_\nu$ for the $\nu$-th band can be evaluated as
\begin{eqnarray}
n_\nu &=& n_{\nu \uparrow}+n_{\nu \downarrow}=\int{d^3{\bf k}\over
(2\pi)^3}\left[1-\frac{\xi_{{\bf k}\nu}}{E_{{\bf
k}\nu}}\Theta(E_{{\bf k}\nu}^\downarrow)\right],\nonumber\\
\delta_\nu &=& n_{\nu \uparrow}-n_{\nu \downarrow}=\int{d^3{\bf
k}\over (2\pi)^3}\Theta(-E_{{\bf k}\nu}^\downarrow),
\end{eqnarray}
and the total density $n$ and total density imbalance $\delta$ of
the system are defined as $n=n_1+n_2$ and
$\delta=\delta_1+\delta_2$.

\subsection {Stability Analysis}
Let us now discuss qualitatively what happens when the mismatch
$h_\nu$ increases. For vanishing mismatch, the system is in a
fully gapped BCS state with $\Delta_1\equiv\Delta_{10}$ and
$\Delta_2\equiv\Delta_{20}$, and the spin density imbalance
$\delta$ is zero. With increasing $h_\nu$, while the BCS state is
still a solution of the gap equations, there may appear another
solution (Sarma) where at least one of the pairing gap
$\Delta_\nu$ is less than the corresponding mismatch, namely
$h_\nu>\Delta_\nu$. In this state, the dispersion of the
quasi-particle $E_{{\bf k}\nu}^{\downarrow}$ becomes gapless and
the system has a nonzero spin density imbalance $\delta$. Note
that the normal state with vanishing condensate is always a
solution of the gap equations and becomes the ground state when
both $h_1$ and $h_2$ are large enough.

Different from the conventional Sarma state in single-band models,
we may have two types of Sarma states in two-band systems. The
first type (type I) is the solution where both mismatches are
larger than the corresponding pairing gaps, namely $h_1>\Delta_1$
and $h_2>\Delta_2$. For this type, there exist gapless excitations
in both bands. The second type (type II) is the solution where
only one mismatch is larger than the corresponding pairing gap,
$h_1>\Delta_1$ and $h_2<\Delta_2$ or $h_1<\Delta_1$ and
$h_2>\Delta_2$. For this type, gapless excitations exist only in
one band. We will show in the following that the stabilities of
these two types of Sarma states are quite different.

\begin{figure}[!htb]
\begin{center}
\includegraphics[width=6.8cm]{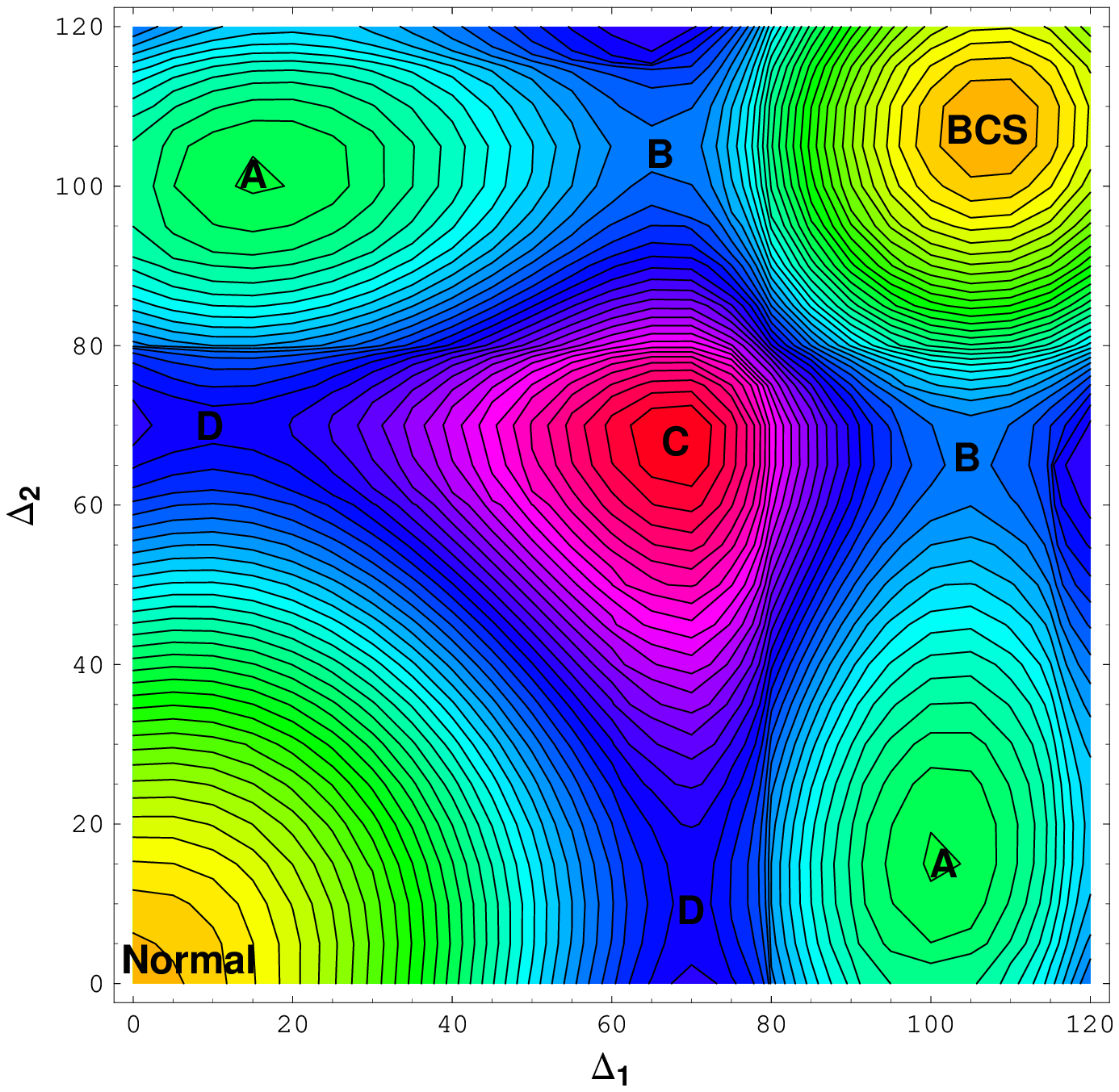}
\includegraphics[width=1cm]{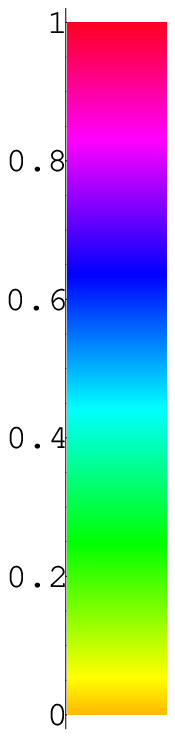}
\caption{The thermodynamic potential contour
$\Omega(\Delta_1,\Delta_2)$ for two symmetric bands with
$U_{1}=U_{2}$. A proper unit is chosen such that the Fermi energy
$\epsilon_{\text F}=200$. The values of the other parameters are
$k_0=100k_{\text F}$, $J=10^{-4}U_0$ with $U_0=4\pi/(mk_{\text
F})$, $(k_{\text F}a_1)^{-1}=(k_{\text F}a_2)^{-1}=-0.5$ and
$h=75$, where $k_{\text F}\simeq\sqrt{2m\mu}$ is the Fermi
momentum. The band on the right shows the relative strength of
$\Omega$ corresponding to different colors. For the parameter
setting, we have $U_1=U_2\simeq0.0156 U_0$, and hence $J\ll
U_1,U_2$. \label{fig1}}
\end{center}
\end{figure}

A numerical example which supports the above argument is shown in
Fig.\ref{fig1} for two symmetric bands with $U_{1}=U_{2}$. For the
sake of simplicity, in our numerical examples presented here, we
assume the same effective masses, chemical potentials and
mismatches for the two bands, i.e., $m_1=m_2\equiv m$,
$\mu_1=\mu_2\equiv\mu$ and $h_1=h_2\equiv h$, this means that only
the total density $n$ and total spin density imbalance $\delta$
can be adjusted. We write $U_1$ and $U_2$ in terms of the s-wave
scattering length $a_\nu$ with a momentum cutoff $k_0$,
$U_{\nu}^{-1}=-m/(4\pi a_\nu)+\int_{|{\bf k}|<k_0} d^3{\bf
k}/(2\pi)^3 m/{\bf k}^2$. Our qualitative conclusions do not
depend on the used regularization scheme. In the case of
$U_1=U_2$, the solutions of the gap equations are distributed
symmetrically in the $\Delta_1-\Delta_2$ plane. Besides the
familiar BCS and normal states which are, respectively, the global
minimum and a local minimum in Fig.\ref{fig1}, we have some Sarma
states in the potential contour. The Sarma states $C$ and $D$ are
of type I, and $C$ is the global maximum and $D$ indicates two
saddle points. The type II Sarma states are marked by $A$ and $B$,
corresponding, respectively, to two local minima and two saddle
points.

If a solution of the gap equations is the ground state of the
system, it should be the global minimum of the thermodynamic
potential $\Omega$~\cite{BP,comment}. The condition for a local
minimum of $\Omega$ is that the matrix
\begin{equation}
{\cal M}=\left(\begin{array}{cc} \frac{\partial^2\Omega(\Delta_1,\Delta_2)}{\partial\Delta_1^2}&\frac{\partial^2\Omega(\Delta_1,\Delta_2)}{\partial\Delta_1\partial\Delta_2}\\
\frac{\partial^2\Omega(\Delta_1,\Delta_2)}{\partial\Delta_2\partial\Delta_1}&
\frac{\partial^2\Omega(\Delta_1,\Delta_2)}{\partial\Delta_2^2}\end{array}\right)
\end{equation}
should have only positive eigenvalues, namely $\det{\cal M}>0$ and
$\text{Tr}{\cal M}>0$. The second order derivatives can be
evaluated as
\begin{eqnarray}
\frac{\partial^2\Omega(\Delta_1,\Delta_2)}{\partial\Delta_\nu^2}&=&\frac{2J}{G}\frac{\Delta_{\bar{\nu}}}{\Delta_\nu}+I_\nu,\nonumber\\
\frac{\partial^2\Omega(\Delta_1,\Delta_2)}{\partial\Delta_\nu\partial\Delta_{\bar{\nu}}}&=&-\frac{2J}{G}
\end{eqnarray}
with the quantities $I_\nu$ defined as
\begin{equation}
I_\nu=\int{d^3{\bf k}\over (2\pi)^3}\frac{\Delta_\nu^2}{E_{{\bf
k}\nu}^2}\left[\frac{\Theta(E_{{\bf k}\nu}^\downarrow)}{E_{{\bf
k}\nu}}-\delta(E_{{\bf k}\nu}^\downarrow)\right].
\end{equation}

For vanishing inter-band coupling $J=0$, the stability condition
becomes
\begin{eqnarray}
I_1I_2>0,\ \ \ \ I_1+I_2>0.
\end{eqnarray}
Note that the properties of the functions $I_1$ and $I_2$ are the
same as the function $I$ defined in the last section. Thus at the
BCS side, namely for $a_1<0$ and $a_2<0$, the Sarma state is
unstable.

Now we discuss how the inter-band coupling $J$ modifies the Sarma
instability at the BCS side. For $J\neq0$, the stability condition
reads
\begin{eqnarray}
&&\frac{2J}{G}\left(\frac{\Delta_1}{\Delta_2}I_1+\frac{\Delta_2}{\Delta_1}I_2\right)+I_1I_2>0,\nonumber\\
&&\frac{2J}{G}\left(\frac{\Delta_1}{\Delta_2}+\frac{\Delta_2}{\Delta_1}\right)+I_1+I_2>0.
\end{eqnarray}
For the type I Sarma state with $h_1>\Delta_1$ and $h_2>\Delta_2$,
we have $I_1<0$ and $I_2<0$. In this case, we can exactly prove
that the above two inequalities can not be satisfied
simultaneously. This type of Sarma state should correspond to the
maximum or saddle point of the thermodynamic potential and is
hence unstable, like the points $C$ and $D$ in Fig.\ref{fig1}.
However, the situation changes for the type II Sarma state.
Without loss of generality, let us discuss the case with
$h_1>\Delta_1$ and $h_2<\Delta_2$. In this case, only the first
band is gapless, and hence $I_1<0$ and $I_2>0$. From $I_2>0$, the
above two inequalities are equivalent to the following one
\begin{equation}
\label{con}
\frac{2J}{G}\frac{\Delta_2}{\Delta_1}+I_1\left(1+\frac{2J}{GI_2}\frac{\Delta_1}{\Delta_2}\right)>0.
\end{equation}
Even though $I_1<0$, this condition can be satisfied, provided
that a nonzero inter-band coupling $J$ is turned on. Suppose the
solution of the gap equations satisfies $\Delta_1\ll\Delta_2$ and
$\Delta_1$ is not quite close to $h_1$, which corresponds to the
case with large polarization $\delta_1$, the first term in
(\ref{con}) is large but the modulus of the second term is
relatively small, and therefore the stability condition can be
satisfied, like the point $A$ at the up-left cornel in
Fig.\ref{fig1}. However, on the other hand, for $\Delta_1\lesssim
h_1$ which corresponds to the case with small polarization
$\delta_1\rightarrow 0$, the absolute value of $I_1$ is very
large, and the Sarma state maybe unstable, which corresponds to
the saddle point $B$ in the upper part of Fig.\ref{fig1}.

We conclude that, in two-band Fermi systems with non-zero
inter-band pairing interaction, the Sarma state can become at
least the local minimum of the thermodynamic potential, and
therefore should be a potential candidate of the ground state.
\begin{figure}[!htb]
\begin{center}
\includegraphics[width=6.5cm]{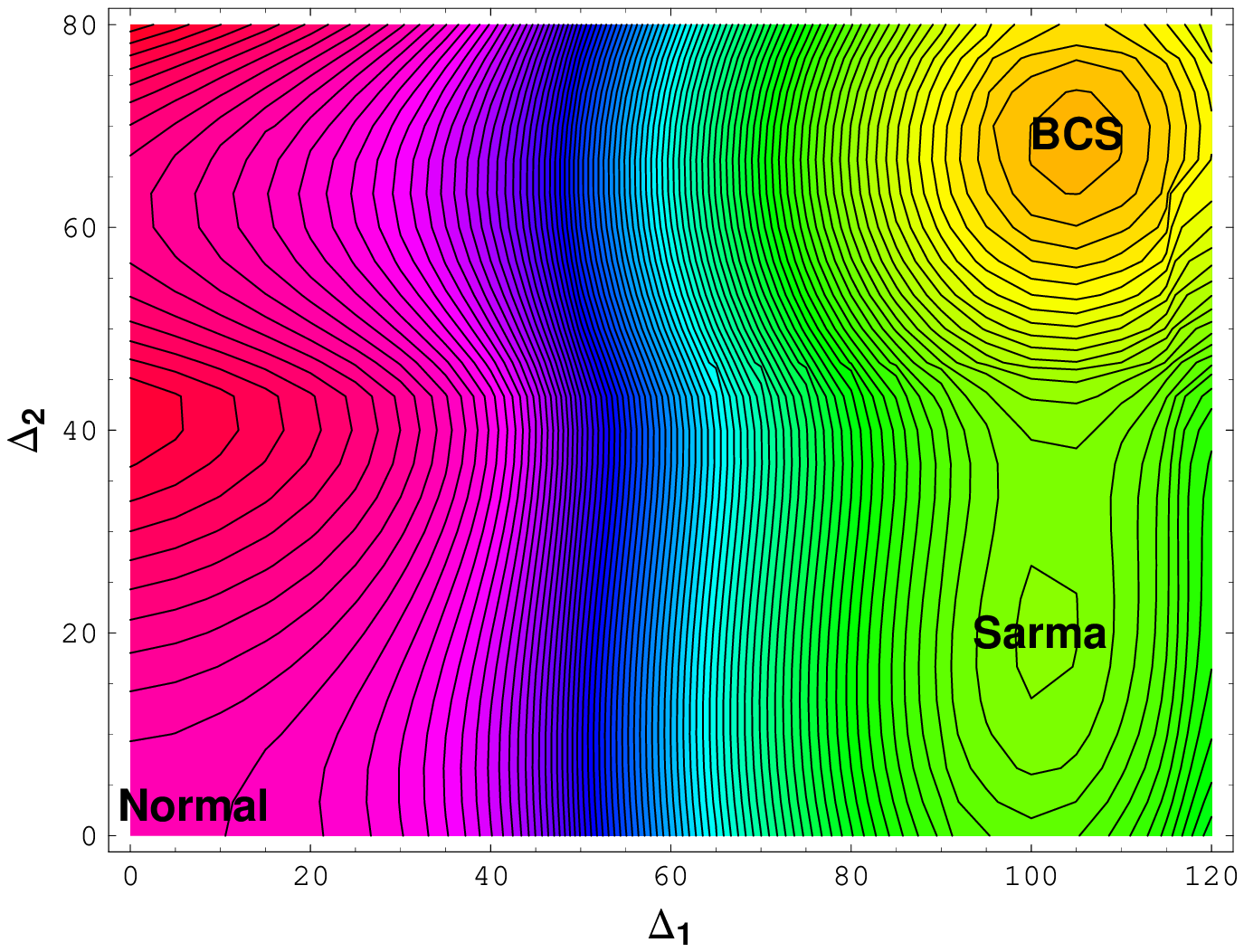}
\includegraphics[width=6.5cm]{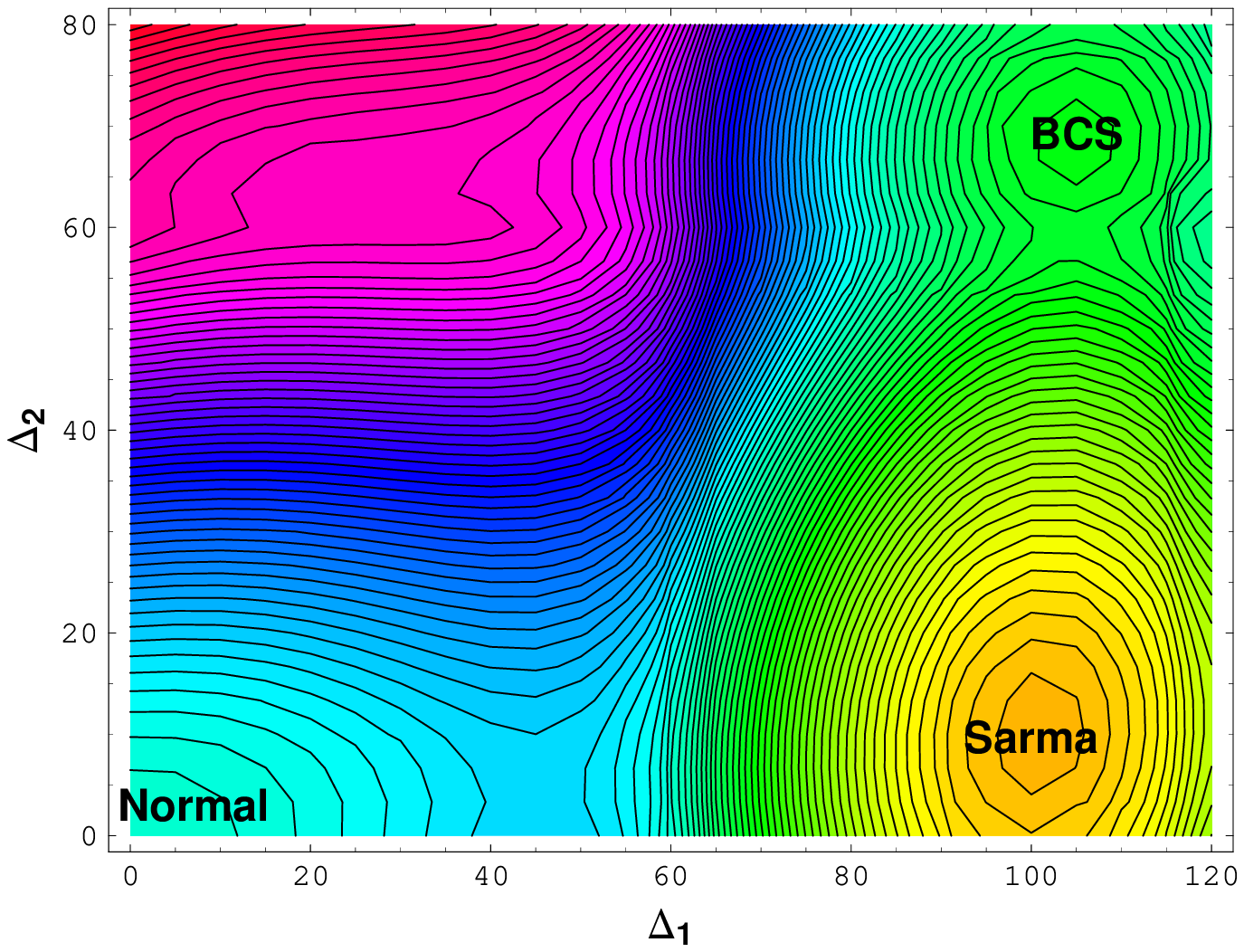}
\includegraphics[width=6.5cm]{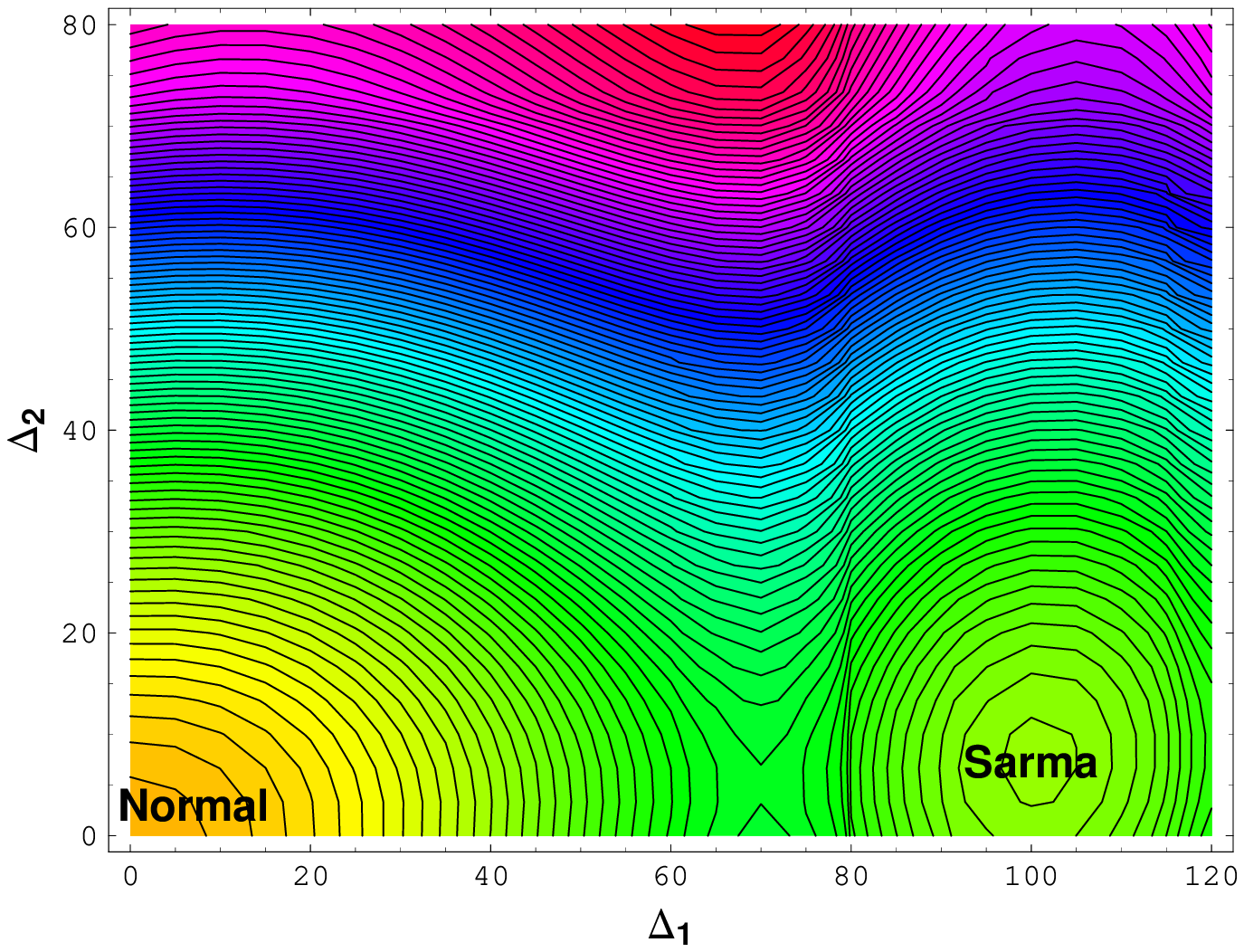}
\caption{The thermodynamic potential contour
$\Omega(\Delta_1,\Delta_2)$ for two different bands with
$(k_{\text F}a_1)^{-1}=-0.5$ and $(k_{\text F}a_2)^{-1}=-0.8$ at
$h=45$ (top), $60$ (middle) and $75$ (bottom). The other
parameters are the same as that in Fig.\ref{fig1}. \label{fig2}}
\end{center}
\end{figure}

\begin{figure}[!htb]
\begin{center}
\includegraphics[width=6.5cm]{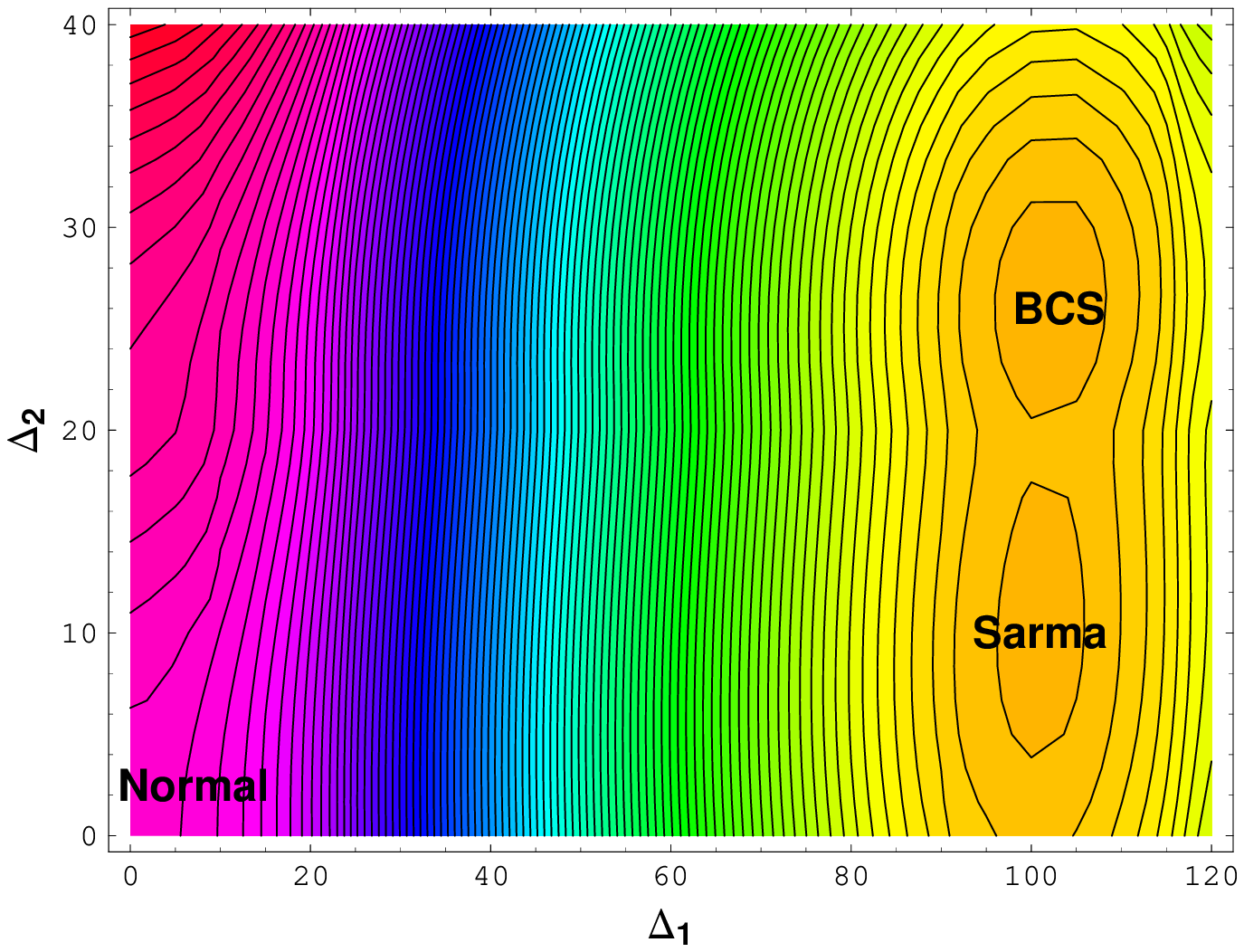}
\includegraphics[width=6.5cm]{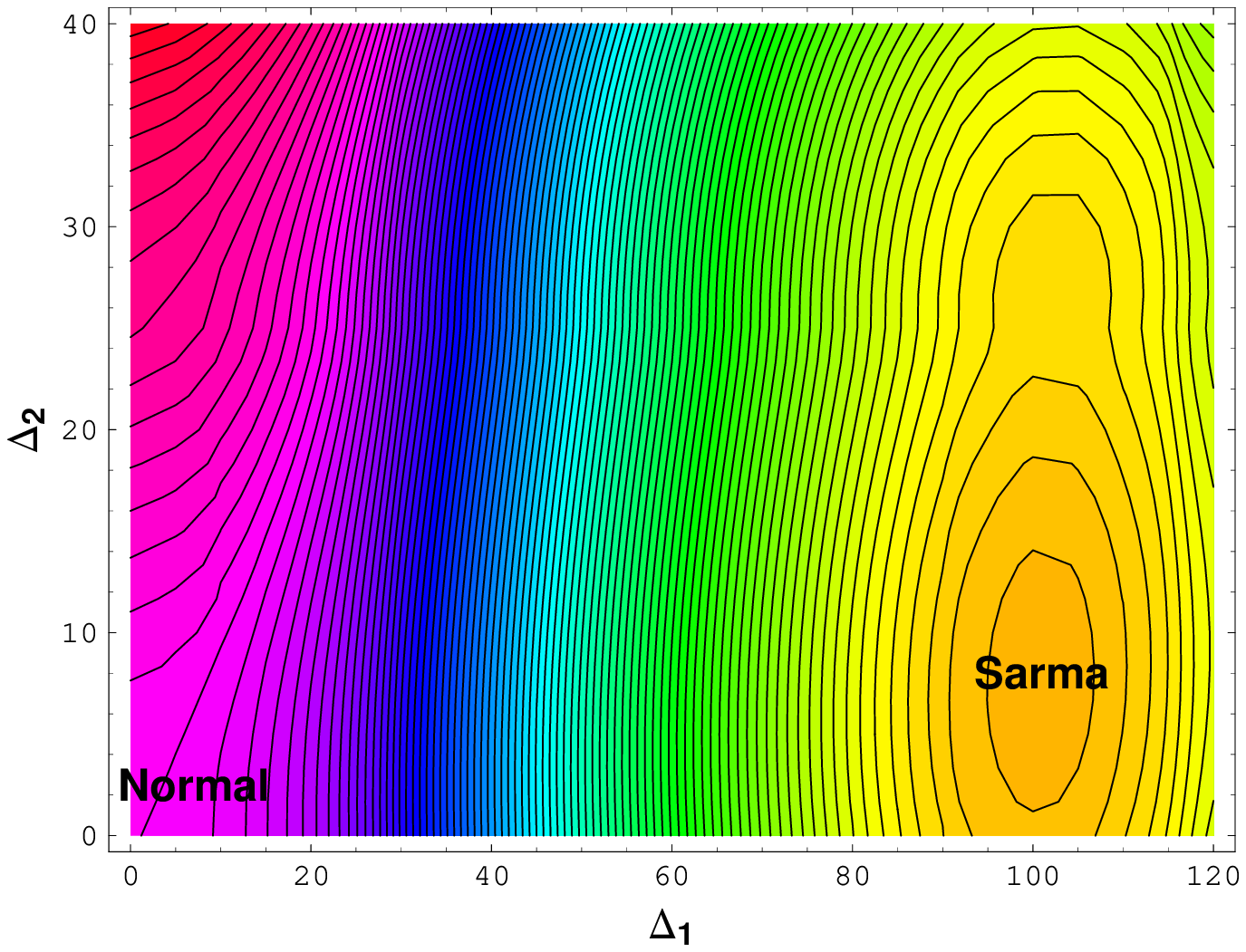}
\includegraphics[width=6.5cm]{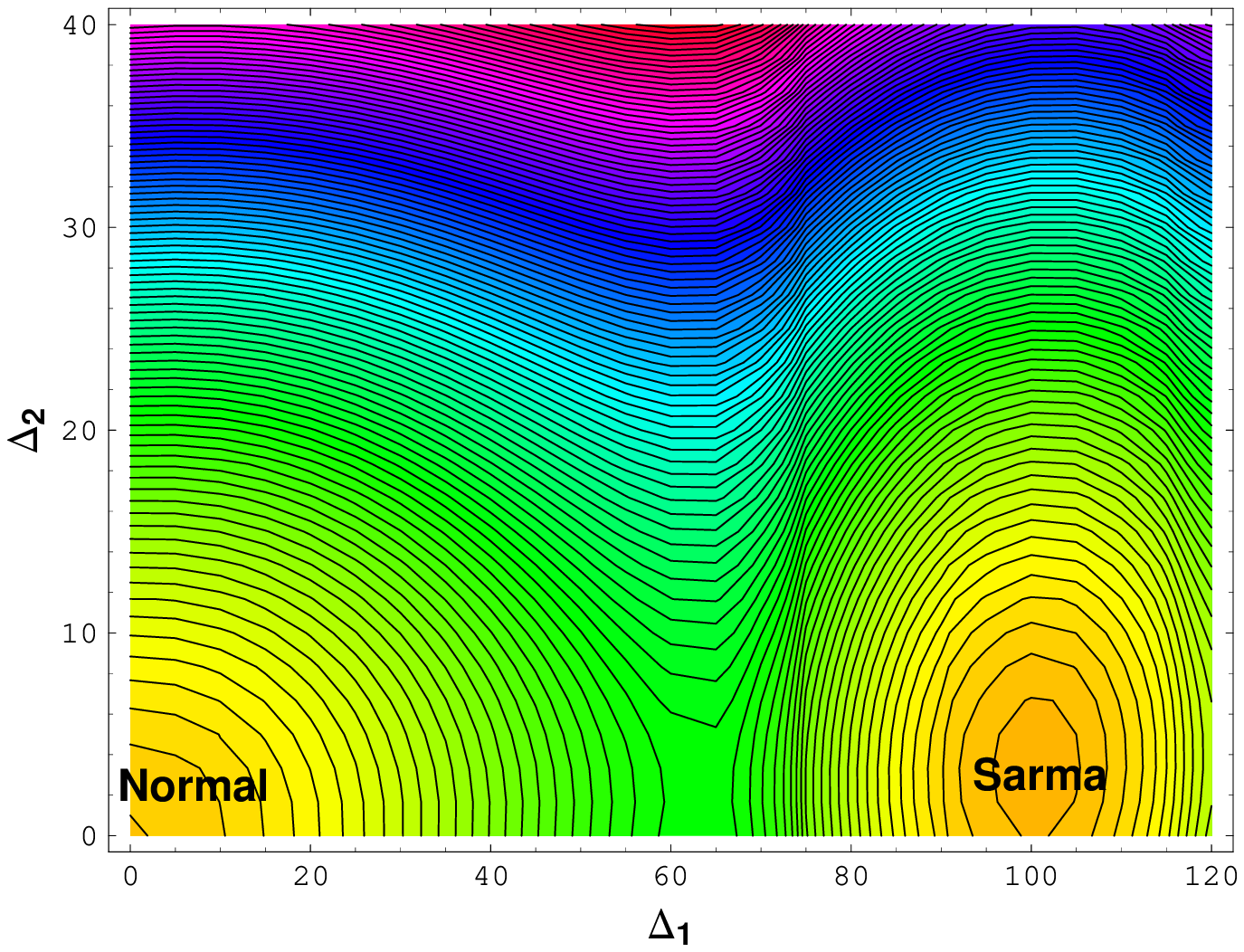}
\caption{The thermodynamic potential contour
$\Omega(\Delta_1,\Delta_2)$ for two different bands with
$(k_{\text F}a_1)^{-1}=-0.5$ and $(k_{\text F}a_2)^{-1}=-1.5$ at
$h=20$ (top), $25$ (middle) and $70$ (bottom). The other
parameters are the same as that in Fig.\ref{fig1}. \label{fig3}}
\end{center}
\end{figure}

However, the condition $J\neq0$ is not sufficient for us to have a
real stable Sarma state. For the case with two symmetric bands
shown in Fig.\ref{fig1}, we found that the global minimum is
always the BCS or normal state for any mismatch $h$, which means
that the Sarma state can not be the ground state even though it
can be a local minimum. However, this can be significantly changed
if some asymmetry between the two bands, such as unequal couplings
$U_{1}\neq U_{2}$, is turned on. In Fig.\ref{fig2} and
Fig.\ref{fig3}, we show the potential contour with $U_{1}\neq
U_{2}$ for three values of $h$. In this case, the number of Sarma
solutions is largely suppressed due to the asymmetry, especially
the state $C$ in Fig.\ref{fig1} as the global maximum of $\Omega$
disappears. Without regard to the saddle points which are
impossible to be stable solutions, the only Sarma state marked in
Fig.\ref{fig2} and Fig.\ref{fig3} appears to be the global minimum
of the system, when the mismatch $h$ is in a suitable region. From
the top to the bottom in Fig.\ref{fig2} and Fig.\ref{fig3}, when
the mismatch $h$ increases, the global minimum changes from the
BCS state to the Sarma state and then to the normal state. In
contrast to the conventional single band model where only one
first order phase transition from the BCS to normal state is
predicted, we have in this two band system two first order phase
transitions when $h$ increases: The first is from the BCS to Sarma
state, and the second is from the Sarma to normal state. The first
order phase transition from BCS to Sarma state was found
in~\cite{BP} by considering the momentum structure of the pairing
gap. For ultracold atom gases, the chemical potential mismatch $h$
should be replaced by the spin population imbalance $\delta$.
However, the phase structure should be essentially independent of
the assembles one used~\cite{BP,comment}, we here do not consider
the case with fixed $\delta$.

Let us compare the numerical results presented in Fig.\ref{fig2}
and Fig.\ref{fig3}. In Fig.\ref{fig3}, the coupling asymmetry is
much larger than that in Fig.\ref{fig2}, we have
$\Delta_{10}/\Delta_{20}\simeq1.5$ in Fig.\ref{fig2} and
$\Delta_{10}/\Delta_{20}\simeq4$ in Fig.\ref{fig3}. We find that
the $h$ window for the Sarma state is wider when the coupling
asymmetry becomes larger. In Fig.\ref{fig2}, the window for Sarma
state is roughly from $h=52$ to $h=71$, and the CC limit is about
$h_c\simeq\Delta_{20}$. In Fig.\ref{fig3}, this window is roughly
from $h=20$ to $h=70$, and the CC limit is about
$h_c\simeq2.7\Delta_{20}$.

\subsection {Solutions at Weak Coupling}
At weak coupling, the same tricks used in Section \ref{s2} can be
employed. For convenience, we define here a function
\begin{eqnarray}
F(\Delta,h)&=&\int_0^\Lambda
d\xi\frac{\Theta(\sqrt{\xi^2+\Delta^2}-h)}{\sqrt{\xi^2+\Delta^2}}\\
&\simeq&\ln\frac{2\Lambda}{\Delta}-\Theta(h-\Delta)\ln\frac{h+\sqrt{h^2-\Delta^2}}{\Delta}\nonumber
\end{eqnarray}
and express the gap equations of our two band model in terms of
it,
\begin{eqnarray}
&&\left[\frac{U_{2}}{GN_1}-F(\Delta_1,h_1)\right]\Delta_1-\frac{J}{GN_1}\Delta_2=0,\nonumber\\
&&\left[\frac{U_{1}}{GN_2}-F(\Delta_2,h_2)\right]\Delta_2-\frac{J}{GN_2}\Delta_1=0,
\end{eqnarray}
where $N_1$ and $N_2$ are the densities of state at the Fermi
surfaces for the two bands. Unlike the single band model, the
above coupled gap equations can not be solved analytically. With
the numerical solutions $\Delta_1$ and $\Delta_2$, the
thermodynamic potential can be evaluated as
\begin{eqnarray}
\Omega=-\sum_{\nu}\left[\frac{N_\nu}{2}\Delta_\nu^2+\Theta(h_\nu-\Delta_\nu)N_\nu
h_\nu\sqrt{h_\nu^2-\Delta_\nu^2}\right].
\end{eqnarray}

\begin{figure}[!htb]
\begin{center}
\includegraphics[width=7.5cm]{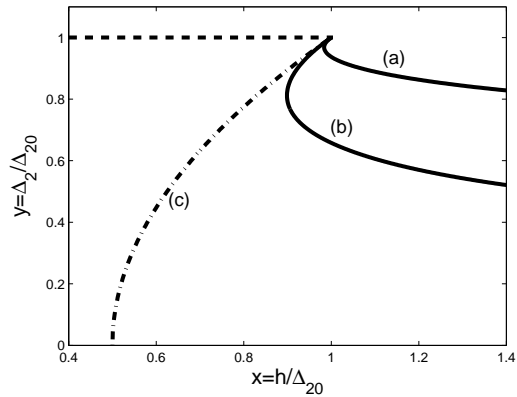}
\caption{The solution $y=\Delta_2/\Delta_{20}$ to the gap
equations as a function of $x=h/\Delta_{20}$. The dashed line
denotes the BCS solution $y=1$ in the region $0<x<1$. The solid
lines (a) and (b) are the Sarma solutions for $J\neq0$. In the
calculations, we take $N_1/N_2=1.5$ and $J/\sqrt{U_1U_2}=0.07$.
For (a) we take $(U_1N_1)/(U_2N_2)=3$ which leads to
$\Delta_{10}/\Delta_{20}\simeq 9$, and for (b) we have
$(U_1N_1)/(U_2N_2)=1.67$ and hence $\Delta_{10}/\Delta_{20}\simeq
3$. The dot-dashed line (c) is the conventional Sarma solution for
$J=0$, $y=\sqrt{2x-1}$ in the region $0.5<x<1$. \label{fig4}}
\end{center}
\end{figure}

For the sake of simplicity, we consider the case $h_1=h_2=h$ which
corresponds to the realistic two-band superconductors in strong
magnetic field. Let us assume $U_1N_1>U_2N_2$ which leads to
$\Delta_1>\Delta_2$. According to the stability analysis, we have
three possible ground states: 1)The normal state with
$\Delta_1=\Delta_2=0$; 2)The gapped BCS state with energy gaps
$\Delta_1\equiv\Delta_{10}>h$ and $\Delta_2\equiv\Delta_{20}>h$;
3)The gapless Sarma state where only $\Delta_1>h$ but
$\Delta_2<h$. We focus here on the case with $\Delta_2\ll\Delta_1$
and $J\ll\sqrt{U_1U_2}$. In this case, the solution of $\Delta_1$
is approximately independent of $h$ and is given by
\begin{equation}
\Delta_1=\Delta_{10}\simeq2\Lambda e^{-1/(U_1N_1)},
\end{equation}
and the Sarma solution for $\Delta_2$ is determined by the
following equation
\begin{equation}
\ln\frac{\Delta_{20}}{h+\sqrt{h^2-\Delta_2^2}}=\frac{J\Delta_{10}}{U_1U_2N_2}\left(\frac{1}{\Delta_{20}}-\frac{1}{\Delta_{2}}\right),
\end{equation}
where $\Delta_{20}$ is obtained by the equation
\begin{equation}
\frac{1}{U_2N_2}-\ln\frac{2\Lambda}{\Delta_{20}}=\frac{J}{U_1U_2N_2}\frac{\Delta_{10}}{\Delta_{20}}.
\end{equation}
We have numerically checked that the above approximation is
sufficiently good for the scaled solution $y=\Delta_2/\Delta_{20}$
as a function of $x=h/\Delta_{20}$. Note that for $J=0$ the
conventional Sarma solution $y=\sqrt{2x-1} (0.5<x<1)$ is
recovered, but for $J\neq0$ the Sarma solution is qualitatively
changed: $y=0$ can not be a solution and there exist solutions for
$x>1$. The solutions for both cases of $J=0$ and $J\neq0$ are
illustrated in Fig.\ref{fig4}. We find that for $J\neq 0$ the
Sarma solution is quite different from the conventional result.
Unlike the well-known Sarma solution which exists in the region
$0.5<x<1$, for $J\neq0$ the Sarma state exists almost in the
region $x>1$ where the BCS solution $y=1$ disappears. Obviously,
in a narrow region $x\lesssim1$ there exists a branch of the
conventional type which is unstable, and the multi-value behavior
of $y$ means a first order BCS-Sarma phase transition at a
critical field $x_1$ which is slightly smaller than $1$.

To discuss the thermodynamic stability of the Sarma state, we then
need to compare it with the normal state. In the case of
$\Delta_2<h$, we find
\begin{eqnarray}
\Omega_{\text{S}}-\Omega_{\text
N}&=&\frac{N_1}{2}(2h^2-\Delta_{1}^2)\nonumber\\
&+&\frac{N_2}{2}\left(2h^2-\Delta_2^2-
2h\sqrt{h^2-\Delta_2^2}\right).
\end{eqnarray}
Some analytical estimations can be made. For large asymmetry
$\Delta_2\ll\Delta_1$, around the $h$-window $h\sim\Delta_{20}$
but $h\ll\Delta_{10}$ for the Sarma state, the sign of the
quantity $\Omega_{\text S}-\Omega_{\text N}$ is dominated by the
first term, if $N_1$ is not much smaller than $N_2$. In this case,
the BCS solution is absent and the Sarma state is the stable
ground state. This argument confirms our conclusion from the
numerical results in Fig.\ref{fig2} and Fig.\ref{fig3}: The $h$
window for stable Sarma state is wider when the asymmetry between
the two bands becomes larger. This means that the CC limit of such
a two-band superconductor can be much higher than the conventional
value $h_c=0.707\Delta_{20}$.

\section {Summary} \label{s4}
We have studied the stability of Sarma state in two-band Fermi
systems via both the stability analysis and analytical solution at
weak coupling. From the stability analysis, the Sarma state can be
the minimum of the thermodynamic potential and hence a possible
candidate of the ground state, if the inter-band exchange
interaction is turned on. Both numerical and analytical studies
show that, a large asymmetry between the two bands or the two
pairing gaps is an important condition for thermodynamic stability
of the Sarma state. When the condition is satisfied, two first
order phase transitions will occur when the mismatch increases,
one is from the BCS to Sarma state at a lower mismatch and the
other is from the Sarma to normal state at a higher mismatch. Our
predictions could be tested in multi-band superconductors and
ultracold atom gases, and such a gapless superconductor may have
many unusual properties, such as magnetism and large spin
susceptibility~\cite{takada,he}.

{\bf Acknowledgments:}\ We thank W.V.Liu and M.Iskin for useful
communications and Y.Liu and X.Hao for the help in numerical
calculations. The work is supported by the NSFC Grants 10575058
and 10735040 and the National Research Program Grant 2006CB921404.

\end{document}